\documentclass[aps,prl,twocolumn, titlepage,showpacs]{revtex4}

\usepackage{graphicx}

\usepackage{dcolumn}

\usepackage{bm}

\begin{document}

\title{Evolution of shear-induced melting in dusty plasma}

\author{Yan Feng}
\email{yan-feng@uiowa.edu}
\author{J. Goree}
\author{Bin Liu}
\affiliation{Department of Physics and Astronomy, The University
of Iowa, Iowa City, Iowa 52242}

\date{\today}

\begin{abstract}

The spatiotemporal development of melting is studied
experimentally in a 2D dusty plasma suspension. Starting with an
ordered lattice, and then suddenly applying localized shear, a
pair of counter-propagating flow regions develop. A transition
between two melting stages is observed before a steady state is
reached. Melting spreads with a front that propagates at the
transverse sound speed. Unexpectedly, coherent longitudinal waves
are excited in the flow region.

\end{abstract}

\narrowtext

\maketitle

Applying shear can induce melting~\cite{Ramaswamy:86, Weider:93,
Butler:02, Delhommelle:04, Messina:06}. Experiments have been
reported in soft materials: colloidal
suspensions~\cite{Ackerson:81, Eisenmann:09}, two-dimensional (2D)
electron crystals with magnetic field ~\cite{Wilen:87},
foams~\cite{Gopal:99}, polymer glasses~\cite{Weitz:09} and dusty
(complex) plasmas~\cite{Nosenko:04}. Most of these studies were
done with a steady application of shear.

If shear were instead applied suddenly, one could investigate the
spatiotemporal development of shear-induced melting. However,
there have been no such experiments reported in any physical
system to our knowledge.

In addition to melting, another result of sudden application of
shear is wave excitation. When strong shear is applied to a 2D
crystal lattice, plastic deformation occurs, and this can cause
melting~\cite{Nosenko:04}. When the applied shear is weaker and
deformation is elastic, transverse waves (phonons) propagate
through the lattice~\cite{Nunomura:00}. Another type of wave is
longitudinal. To our knowledge, at least within the literature for
dusty plasmas, there have been no reports of the excitation of
coherent {\it longitudinal} waves due to the application of {\it
shear} in either the elastic or plastic regimes.

Here we seek answers to three questions. First, can coherent
longitudinal waves be generated by applying shear? Second, when
shear is applied suddenly, and melting occurs, what is the
spatiotemporal development? Third, is there a melting front, and
how does it spread?

Physical systems that allow motion essentially on a 2D plane
include a Wigner lattice of electrons on a liquid-He
surface~\cite{Wilen:87}, ions confined in a Penning
trap~\cite{Mitchell:99}, colloidal suspensions~\cite{Murray:90},
granular fluids~\cite{Reis:06}, vortex arrays in the mixed state
of type-II superconductors~\cite{Gammel:87}, and dusty plasmas
levitated in a single layer~\cite{Feng:08, Ratynskaia:06}.

Melting of 2D systems has a different mechanism than for 3D
systems~\cite{Murray:90}; it is sometimes termed an order-disorder
transition~\cite{Reichhardt:08}. Order and disorder are
characterized by measures of structure. For example, the local
six-fold bond-orientational order $|\phi_6|$~\cite{Delhommelle:04,
Weider:93, Woon:04} measures local order, while global order can
be characterized by the abundance of defects~\cite{Feng:08,
Reichhardt:08}, which are non-hexagonal Wigner-Seitz cells in
Voronoi diagrams~\cite{Delhommelle:04, Feng:08}.

Dusty plasma is partially ionized gas containing micron-size
particles of solid matter~\cite{Feng:08}. Dusty plasmas allow
atomistic scale observation of dynamics by tracking particles with
video microscopy, and they also allow laser manipulation of
particles~\cite{Feng:08}. We will exploit these capabilities to
observe at an atomistic scale, with both spatial and temporal
resolution, the sudden onset of shear-induced melting. In most
previous dusty plasma experiments, melting was studied under {\it
steady-state} conditions (by changing plasma
parameters~\cite{Thomas:96, Melzer:96, Sheridan:08}, varying
particle number~\cite{Ivlev:03}, and laser
manipulation~\cite{Nosenko:04, Nosenko:06}). The {\it temporal
development} of melting has received less study, and experiments
that have been reported~\cite{Samsonov:04, Feng:08} relied on
mechanisms other than shear-induced melting.

Particles have an electrical charge $Q$ and are electrically
confined in a single horizontal layer where they self-organize
with a structure analogous to a crystalline solid or a
liquid~\cite{Feng:08}. Coulomb repulsion is shielded with a
screening length $\lambda_D$ ~\cite{Konopka:00}. Dusty plasmas are
driven-dissipative systems~\cite{Liu:08, Feng:08}, with frictional
drag on the rarefied gas at a damping rate $\nu_f$~\cite{Liu:03}.
The collective oscillation of particles can be characterized by
the nominal 2D dust plasma frequency
$\omega_{pd}$~\cite{Kalman:04}.

For a dusty plasma suspension, shear must be applied differently
than for most substances (e.g., colloidal
suspensions~\cite{Eisenmann:09}) because the suspension does not
contact a container. We apply shear internally within our sample,
using laser radiation pressure~\cite{Nosenko:04, Woon:04}.

Using the apparatus of \cite{Liu:08}, an Argon plasma was
generated in the vacuum chamber at $15.5~\rm{mTorr}$, powered by
$13.56~\rm{MHz}$ radio-frequency voltages at $184~\rm{V}$ peak to
peak. The particles were polymer microspheres with a diameter
$8.1~\rm{\mu m}$ and a gas damping rate
$\nu_f=2.7~\rm{s^{-1}}$~\cite{Liu:03}.

The particles were suspended in a single layer. Before applying
shear they self-organized in a triangular lattice with six-fold
symmetry~\cite{Feng:08}. Particle motion was essentially 2D, with
negligible out-of-plane displacements and no buckling of the
particle layer. The suspension had a diameter $\approx 52~\rm{mm}$
and contained $> 11~000$ particles. About 2800 particles were in
the analyzed region, $(24.7 \times 20.4)~\rm{mm^2}$. The particle
spacing was characterized by a Wigner-Seitz
radius~\cite{Kalman:04} $a = 0.25~\rm{mm}$.

To apply shear stress with a sudden onset, we used laser radiation
pressure, which applies forces internally within the
sample~\cite{Nosenko:04}. The power of a pair of 532-nm laser
beams was increased within $40~\rm{ms}$ to a constant level. A
different constant level was chosen for each of our four runs,
varying from 0.57 to $1.90~\rm{W}$ per beam, as measured inside
the chamber. The beams pointed oppositely in the $\pm x$
directions and struck the suspension at a 6$^\circ$ downward
angle. Their width and separation were $\Delta Y_l \approx
0.2~\rm{mm}$ and $L = 4.7~\rm{mm}$, respectively. To apply shear
across the entire suspension width, the beams were rastered across
the full suspension (at a frequency high enough, $200~\rm{Hz}$, to
avoid exciting coherent waves at the rastering
frequency~\cite{Nosenko:06}).

Using a top-view camera, we recorded videos for a duration of
$14~\rm{s}$ (including $5~\rm{s}$ before applying shear). While
applying shear, particles flowed out of the camera's field of view
(FOV); they circulated around the suspension's perimenter and then
reentered the FOV~\cite{Nosenko:04}. Between runs we turned laser
manipulation off for $\approx$ 20 min, so that the suspension had
enough time to cool, solidify and anneal~\cite{Feng:08}. Then we
calculated particle positions and velocities in each video
frame~\cite{Feng:08}.

Using the wave-spectra analysis method for particle random motion
in an ordered lattice~\cite{Nunomura:02}, we found these
parameters for our dusty plasma: $\omega_{pd}=72.6~\rm{s^{-1}}$,
$Q/e=-(8700\pm900)$, $\lambda_D=(0.33\pm0.07)~{\rm{mm}}$. The same
method also yielded the transverse and longitudinal sound speeds,
$C_t=(4.0\pm0.4)~\rm{mm/s}$ and $C_l=(15.0\pm1.5)~\rm{mm/s}$,
respectively. When coherent waves are observed experimentally,
they can be identified as transverse or longitudinal waves by
their speed and by the direction $v_x$ or $v_y$ of particle
motion.

The experiment was designed to have symmetry with an ignorable
coordinate $x$. We average some quantities over $x$, as denoted by
$\langle \rangle_x$ (averaging with cloud-in-cell weighting with a
bin width of the lattice constant, $1.9~a$). The kinetic energy
(KE) is averaged over both $x$ and $y$ for the full analyzed
region, as denoted by $\langle \rangle_{xy}$. We calculate KE =
KE$_x$ + KE$_y$, where \rm{KE}$_x = m\langle
v_x^{2}\rangle_{xy}/2$, and similarly for \rm{KE}$_y$. Although
this definition includes the energy of directed flow as well as
random motion, we will report the values in temperature units
($K$), to allow comparison to other experiments.

Results in Fig. 1(a),(b) show the development of KE and the flow
velocity after we suddenly applied shear stress. The KE increases
dramatically, by about two orders of magnitude, Fig.~1(a). Melting
occurs for laser powers above a threshold $\approx~0.4$~W. The
flow pattern, Fig.~1(b), broadens until reaching a steady state at
$\approx$ 1 s, when the full width at half maximum (FWHM) of the
velocity profile is $\Delta Y_f=1.2~\rm{mm}$, which we will refer
to as the flow regions. This width $\Delta Y_f$ is determined by a
combination of shear stress applied by the laser, gas friction,
and shear viscosity of the suspension~\cite{Nosenko:04}. The KE is
greater for velocities in the $x$ direction than in the $y$
direction because shear was applied in the $x$ direction ($\approx
53\%$ of \rm{KE}$_x$ is due to directed flow in the steady state).
Scale lengths in the flow are ordered as $\Delta Y_l \approx a <
\Delta Y_f < L$. The flow is laminar, with a Reynolds number $\ll
10^2$~\cite{Nosenko:04}. In the flow region, the kinetic
temperature (estimated from $v_y$) is high, with $\Gamma \approx
20$; this is well beyond the predicted melting point
$\Gamma=160$~\cite{Hartmann:05}.

To answer our first question, we observe coherent {\it
longitudinal} waves propagating away from the flow region. This
result might be surprising because the energy input was purely
shear. These longitudinal waves are revealed in Fig.~1(c) by
wavefronts with a slope corresponding to the longitudinal sound
speed, $C_l$. The wavefronts emerge from the flow region,
suggesting that the longitudinal waves were generated there. These
wavefronts repeat, indicating the waves are coherent, not random.

The coherent longitudinal waves in Fig.~1(c) have a period
$\approx 4\pi\omega_{pd}^{-1}$. This is the almost the same as the
period where the dispersion relations of longitudinal and
transverse waves cross~\cite{Nunomura:05}. This observation
suggests that the coherent longitudinal waves we observe could be
explained by scattering from transverse waves.

Energy is carried into the surrounding lattice by these coherent
longitudinal waves. We will consider the speed of these waves
below, when explaining sudden shear-induced melting.

To answer our second question, we find that after sudden
application of shear, melting occurs in two steps. Examining
Fig.~2(a), we can see that defects initially proliferate at a
higher rate, then soon after at a slower rate, with a distinctive
transition between. We term these ``melting stages'' 1 and 2.
Afterwards, at a long time, there is of course a steady state. We
confirmed that the transition is ubiquitous: it occurs for all
four laser powers we tested, and also in other suspensions with
various values of the particle spacing.

Our experiment provides both spatial and temporal resolution,
which we can exploit to investigate the cause of the transition
between the two melting stages. We are motivated by Fig.~1(b) to
analyze time series separately in two spatial portions: inside and
outside the flow regions.  This yields the time series in
Fig.~2(b), where the quantity plotted is the defect number
fraction, calculated by dividing the number of defects by the
number of particles.

Figure~2(b) reveals that defects proliferate in different places
at different times. In melting stage~1, they proliferate mainly
inside the flow regions until saturating at $t = 0.16~{\rm{s}}$.
Importantly, this is the same time as the transition in Fig.~2(a).
Only afterwards do defects spread widely outside the flow region,
as seen in the bottom curve in Fig.~2(b). Thus, the transition in
the phase diagram arises from the different timing of when defects
appear in the two different regions: first inside the flow region
until defects saturate there, and then outside.

We can quantify the physical time scales for the two melting
stages. Candidate time scales include $\omega_{pd}^{-1}$, which
characterizes mutual particle motion due to interparticle electric
fields, and $\nu_f^{-1}$ due to gas friction. Initially, when
velocities are low, we expect friction to be small, leading us to
consider $\omega_{pd}^{-1}$ as the time scale for melting stage~1.
We find that melting stage~1 has a duration on the order of
$10\,\omega_{pd}^{-1}$, no matter how much shear is applied, as
shown in Fig.~2(a). Later, as the KE increases, Fig.~2(b),
dissipation by gas friction grows until it balances the energy
input from the laser, yielding a steady state. The role of
friction in limiting the spread of melting is suggested by noting
that the duration of melting stage~2 is $\approx 1.5~\nu_f^{-1}$.

To answer the third question, we find that there is a distinctive
melting front, and it propagates at about the transverse sound
speed, $C_t$. This is seen in Fig.~3, where the colored contours
show how disorder spreads with time. We also draw lines with a
slope corresponding to $C_t$; comparing these to the melting
front, we see that they coincide. Thus, the melting front
propagates at about $C_t$, not the much faster $C_l$. We verified
this result using three different measures of structure. These
include the local six-fold bond-orientational order $|\phi_6|$~
\cite{Delhommelle:04, Weider:93, Woon:04} (as shown in Fig.~3),
defect number fraction, and height of the first peak of the local
pair correlation function.

Before this experiment, it was not obvious whether the melting
front should spread at a rate corresponding to $C_t$, $C_l$, or
thermal conduction. Thermal motion can be decomposed into
incoherent waves that include both transverse and longitudinal
modes. Coherent waves are also present, including the surprising
longitudinal modes we found propagating away from the flow region.
Thermal conduction due to temperature gradients~\cite{Fortov:07,
Nosenko:08, Hou:09} would also lead to a spreading of energy as
time passes.

To determine whether the flow of energy is in the form of waves or
thermal conduction~\cite{Fortov:07, Nosenko:08, Hou:09}, we varied
the laser power, which will vary the gradients. Linear waves
propagate at the same speed, regardless of their amplitude; but
energy spreads by thermal conduction more rapidly if the gradients
are larger. We found that the melting front propagated at nearly
$C_t$ for all four laser powers we tested. Thus, we dismiss
thermal conduction as the main mechanism for the propagation of
energy that results in the melting front.

Our result that the melting front propagates at about $C_t$
suggests that the propagation of transverse waves is the mechanism
for transferring the energy required for melting in our
experiment. Transverse waves, which were excited when shear was
first applied, propagate outward carrying energy that can create
defects. As this wavefront travels, it loses energy by creating
disorder and by gas friction. Eventually, after traveling about
3~mm, the outward traveling wave loses so much energy that it can
no longer generate more defects, and the steady state is reached.
It is possible that in other systems, with greater applied shear,
the generated coherent longitudinal waves might be strong enough
to melt a solid lattice, so that the melting front would propagate
at $C_l$.

In summary, we found that coherent longitudinal waves are excited
in this shear-induced melting system. Applying shear suddenly led
to melting in two stages separated by a distinctive transition.
After defects saturated within narrow flow regions, they spread
wider with a melting front that propagates at about $C_t$.

This work was supported by NSF and NASA.

\begin{figure}[p]
\caption{\label{condition} (color online). (a) Time series of the
average kinetic energy (KE). Shear was applied suddenly by
applying 0.95W of laser power starting at $t = 0$. The KE rose
about two orders of magnitude until a steady state was reached at
$\approx$ $1~\rm{s}$. (b) Spatiotemporal evolution of the flow
velocity $\langle v_x \rangle_x$. Two counter-propagating flow
regions broaden to a width $\Delta Y_f$ (defined as FWHM of the
velocity profile). Length scales, including the thickness $\Delta
Y_l$ and spacing $L$ of the laser sheets, are ordered as $\Delta
Y_l \approx a < \Delta Y_f < L$. (c) Coherent longitudinal waves,
generated near the flow region and propagating outward. This is
revealed by the repeated pattern of wavefronts in this plot of
$\langle v_y \rangle_x$. For comparison, the longitudinal sound
speed $C_l$ is indicated by the straight lines. The observed wave
period is $\approx 4\pi\omega_{pd}^{-1}$.}
\end{figure}

\begin{figure}[p]
\caption{\label{stages} (color online). (a) Phase diagram, showing
the proliferation of defects as the KE increases, recorded at
$0.018~\rm{s}$ intervals, for three laser powers. A transition
between two stages is revealed. The quantity plotted is the defect
area fraction; this measure of structure is the area of a Voronoi
diagram that is occupied by defects~\cite{Feng:08}, divided by the
total area. (b) Time series for KE and defects, counted separately
inside and outside the flow regions of width $\Delta Y_f$. The
timing indicates that defects proliferate outside only after they
saturate inside the flow regions. (The inset shows about 1/4 of
the Voronoi diagram, for the run at 0.95 W at $t=0.16~\rm{s}$.)}
\end{figure}

\begin{figure}[p]
\caption{\label{structure} (color online). Spatiotemporal
evolution of orientational order, $|\phi_6|$, which is defined to
have a maximum of unity for a perfectly crystalline
region~\cite{Delhommelle:04, Weider:93, Woon:04}. Lines are drawn
with the slopes $C_t$ and $C_l$, starting at $t = 0$. The
transverse sound speed, $C_t$, coincides with the melting front
propagation. The resolution of our data along the $Y$ axis is
$1.9~a$.}
\end{figure}

\end{document}